\documentclass{aastex631}

\usepackage{booktabs} 

\shorttitle{And if Vulcan was a primordial black hole of planetary-mass ?}
\shortauthors{Pogossian}
\begin{document}

\title{And if Vulcan was a primordial black hole of planetary-mass?}

%%email{pogossia@univ-brest.fr}

\author{Souren P. Pogossian}
\affiliation{Univ. Brest, CNRS, IRD, Ifremer, IUEM, Laboratoire d'Oc\'eanographie Physique et Spatiale (LOPS),\\Technopôle Brest-Iroise, Rue Dumont d'Urville, 29280 Plouzan\'e, Universit\'e de Bretagne Occidentale \\Brest, France\\}

\footnote{pogossia@univ-brest.fr}

\begin{abstract}

In this work, I re-examine the question of a possible explanation for the anomalous advance of Mercury's perihelion by the existence of a hypothetical planet, Vulcan, which I consider to be a kind of primordial black hole of planetary-mass. The detection of this kind of celestial body has become possible with modern experimental techniques, inter alia, such as the Optical Gravitational Lensing Experiment. Recently, an excess of ultra-short microlensing events with crossing times of 0.1 to 0.3 days has been reported, suggesting the possible existence of sub-Earth-mass primordial black holes in our solar system. The primordial black hole Vulcan planetary mass hypothesis could then explain the anomalous advance of Mercury's perihelion under the influence of its gravitational attraction, still remaining hidden from astronomers' telescopes. But in this case, it will also influence the perihelion advance of the other planets. To this end, I first calculate the mutual partial contributions to the perihelion motion of all the planets by two different methods without Vulcan in a model of simplified solar system consisting of the Sun and eight planets. 
Next, I include Vulcan in this model within the framework of the Newtonian theory of classical gravitation and analyze Vulcan's influence on the perihelion advance of the inner planets, using Vulcan parameters from my previous work.  These results are compared with the perihelion advances of the inner planets predicted by the theory of general relativity, and with the data obtained by modern observations.

\end{abstract}


\section{Introduction} \label{sec:intro}

The orbit of Mercury is of constant interest because it is currently accepted that the anomalous advance of Mercury's perihelion can only be explained on the basis of general relativity (\cite{Clemence1947}, \cite{Rana1987}, \cite{Stump1988}, \cite{Morrison1975}, and \cite{Smulsky2011}). Especially since the advance of Mercury's perihelion is considered one of the keystones proving Einstein's theory of General Relativity (GR).
	
The anomalous advance of Mercury's perihelion  is  small of the order of $43^{\prime\prime}$/cy while the biannual fluctuations of Mercury's perihelion advance can be as large as $27^{\prime\prime}$ (\cite{Narlikar1985}, \cite{Pogossian2022}). Given the somewhat ambiguous calculations of this last quantity, I made a detailed study with three different definitions of the perihelion advance (PA), but my calculations only confirmed the value of the PA of about $532^{\prime\prime}$/cy based on the Newtonian classical theory of gravitation, instead of the $575^{\prime\prime}$/cy observed by very recent and refined measurements of \cite{Park2021}. For a more detailed description and some references see \cite{Pogossian2022}.
	
Nineteenth-century scientists were aware of this discrepancy. In order to account for this inconsistency of the Newtonian classical theory of gravitation, \cite{LeVerrier1859} proposed in 1859 the existence of another hypothetical planet inside the orbit of Mercury, which he named Vulcan and whose gravitational influence on Mercury would explain the observed anomalous PA. Le Verrier had already succeeded in predicting the existence of Neptune on a purely theoretical basis which successfully explained the perturbations of the orbital motion of Uranus. The Vulcan hypothesis proposed by Le Verrier poses a fundamental challenge: either the observations confirm the existence of Vulcan, or one must admit the failure of Newton's classical theory of gravitation.
	
A real hunt for the observation of this planet began with the first record of Lescarbault (\cite{Baum1997}), who in 1859 observed the passage of a black disk on the background of the Sun with sharp contours.  After a few years of inconclusive research by probing the intra-mercurian space by telescopes, astronomers began to be puzzled because Lescarbault's observations were not confirmed.  
	
Le Verrier's initial hypothesis was the existence of a single massive planet between Mercury and the Sun that he hoped to observe during the transit of Vulcan. But the long absence of observations of a Vulcan transit led him to look for another possible form of the Vulcan hypothesis, that of an intra-mercurian asteroid belt of fairly large total mass. This is probably the reason why many astronomers have long looked for a belt of vulcanoids rather than a single massive planet.
	
Let compare the orders of magnitude of the population linear density of the main asteroid belt between Jupiter and Mars to the hypothetical vulcanoid belt. The main asteroid belt is only about 1/137.5 of the mass of the Mercury (\cite{Pitjeva2018}) with a mean distance from sun of about $\sim$ 2.66 AU.  If one admits that the total mass of Vulcan is distributed over numerous asteroids, the following considerations can be made on the basis of my previous work concerning this hypothetical planet (\cite{Pogossian2023}). According to the latter reference, the smallest optimized mass for Vulcan giving an experimentally sound value of APM was calculated to be 1.6 masses of Mercury and therefore 220 times larger the total masse of main asteroid belt. Moreover, taking into account the average circumference of the main asteroid belt ($\sim$2.66 AU) and that of the hypothetical vulcanoid belt ($\sim$0.2 AU), it can be concluded that the mass-to-circumference ratio of the vulcanoid belt region should be at least 2926 times greater than that of the main asteroid belt. The presence of bodies of such high linear density would certainly not have escaped observations for so long. Thus, the existence of a large number of intra-mercurian asteroids is made virtually impossible by all observations made so far (\cite{Durda2000}, \cite{Stern2000}, \cite{Steffl2013}, and \cite{Zhao2009}).  In an earlier work, based on numerous observations in the vulcanoid region and on the impossibility of observing Vulcan either as a single planet or as a vulcanoid belt, I showed the invalidity of the Vulcan hypothesis to explain the anomalous advance of Mercury's perihelion (\cite{Pogossian2023}).  
	
Yet the Vulcan hypothesis may still have a chance. Recent theoretical developments point to another possibility for the existence of planetary mass in intra-mercurial space in the form of very dense matter, which could be a Primordial Black Hole (PBH).
	
What if the hypothetical planet Vulcan were not a typical planet or a vulcanoid belt, but a PBH instead? With a mass 1.6 times that of Mercury, Vulcan’s Schwarzschild radius would be about 0.7 mm; at 3.4 times Mercury's mass, it would be around 1.7 mm.
	
As shown in the following references [\cite{Carr2021}, \cite{Krasnov2023}, \cite{Miller2021}, and \cite{Montero-Camacho2019}] hypothetical PBHs are thought to have been born at the time of the Big Bang, and their populations are scattered throughout the cosmos. We could also have a number of PBHs in our solar systems in the form of small planets that are virtually invisible to astronomers [\cite{Barco2021}].
	
Assessing the probability of a PBH being captured by our solar system is challenging, as it relies on various cosmological models and the mechanisms through which PBHs can form (\cite{Carr2021}).  \cite{Eroshenko2023}, based on the work of \cite{Napier2021}, estimated the probability of a PBH being captured by an already formed solar system to be approximately $\sim 4 \times 10^{-8} $ f, f is the fraction of the PBHs and dark matter. Additionally, based on  the work of \cite{Lingam2018}, Eroshenko suggest a higher probability, around $\sim  10^{-4}-10^{-3}$, for capturing a planetary-mass PBH. \cite{Scholtz2020} advanced idea that the 9-th hypothetical planet of solar system may be a PBH.  Thus, these recent studies suggest that, while the probability of capture is low, it remains a viable scenario within our solar system, warranting further investigation.
	
For the detection of such objects different observational methods were proposed very recently.  \cite{Miller2021} suggested detecting planetary-mass PBHs by gravitational waves while \cite{Herman2021} with resonant electromagnetic cavities based on TE, TM waveguides under strong magnetic fields. \cite{Niikura2019} claim microlensing events that  give hint of PBHs existence.  \cite{Domenech2022} have shown that stochastic background of gravitational waves detected by  American Nanohertz Observatory for Gravitational Waves,  naturally predicts substantial existence of planet-mass PBHs. \cite{Arbey2020} suggest that a fly-by spacecraft might even measure the Hawking radiation of the PBH in the radio band. Looking further to the future, gravitational-wave observatories in space might detect the dynamical effects of PBHs.  With such an impressive variety of modern experimental methods at our disposal, we can remain optimistic that Vulcan could be detected if it were a PBH.
	
Throughout this article, I'll consider Vulcan to be a kind of planetary mass PBH whose very small size makes it invisible to observations with optical instruments. Note that in this hypothesis, the density considerations play only a minor role since it is enough to admit the density of Vulcan is so high that it’s geometrical small size makes Vulcan unobservable by conventional observation methods. 
	
Thus this planetary mass  PBH Vulcan hypothesis may explain the anomalous advance of Mercuries perihelion, remaining hidden from the telescopes of astronomers.  But if such a Vulcan exists, its influence on other planets, especially on closest to it (Venus, Earth and Mars), would be detectable. 
	
In this work, I calculate the orbital perturbations of Venus, Earth and Mars in the presence of a Vulcan of sufficiently large mass to induce a PA of $43^{\prime\prime}$/cy on Mercury's orbit. In particular, I calculate the PA of Venus, Earth and Mars with-and-without Vulcan, in the frame of classical Newtonian gravitational theory.  Two sets of parameters for Vulcan will be considered, the first having the orbital parameters reported by Le Verrier but with appropriate mass ($\sim$ 3.37612 Mercury mass) to assure an anomalous advance for Mercury of about $43^{\prime\prime}$/cy   and the second set including optimized parameters calculated in my previous work on Vulcan  (with a Vulcan of about ~ 1.6 Mercury mass) (\cite{Pogossian2023}).
	
Like models based on general relativity (GR), the Vulcan hypothesis offers a possible explanation for the anomalous perihelion advances of planets that cannot be fully explained by classical Newtonian gravitation without including Vulcan. The independent prediction of anomalous perihelion advance by the Vulcan approach and GR-based calculations underlines the importance of the Vulcan model.
	
This study therefore examines the relevance of the Vulcan hypothesis. It also analyzes the influence of each planet on the perihelion motion of the inner planets, uncovering key features that deepen our understanding of planetary dynamics.

\section{Calculations}
\label{sec:2}
The calculation method has been described in detail in my previous publications (\cite{Pogossian2022}, \cite{Pogossian2023}). In these calculations, I used initial conditions in the framework of classical Newtonian gravitation (\cite{Leguyader1993}, \cite{Arminjon2004}, \cite{Standish1990}), without taking into account perturbations  predicted by GR. This model takes into account only 8 planets (and 9 planets when Vulcan is included) in translation around the sun. The Newtonian gravitational equations for a 9-body problem (10-body problem  when Vulcan is included) were integrated over a time interval of 333473 days ($\sim$913 years), with a time step of around 60 minutes. Jupiter and Saturn are the most massive planets and the closest to the inner planets, and are therefore the main source of perturbations to their quasi-elliptical orbits. The orbits of these massive planets are locked by a mean motion resonance of 2:5, and about every 900 years the direction of alignment of these giant planets returns at almost exactly its original direction (\cite{Wilson1985}, \cite{Pogossian2023}). I have therefore chosen a time interval close to the period of "great inequality" to take into account all the possible relative positions of these giant planets with respect to the orbit of each of the inner planets.  As described in detail in my previous works, the 9-body solar system (10-body problem  when the Sun and Vulcan are included)  has been integrated on the basis of purely Newtonian initial conditions taken from DE200/LE200 ephemerides, subtracting corrections based on GR (\cite{Leguyader1993}, \cite{Arminjon2004}).	

The calculation performed by the MATLAB ODE113 solver has been described in detail in my previous works (\cite{Pogossian2022}, \cite{Pogossian2023}). For simplicity I will use only the Laplace Runge Lentz (LRL) vecteur rotation as a measure of the PA of inner planets. In a previous work (\cite{Pogossian2022}) I have shown that among three possible definitions of the PA, the description of Mercury’s PA by the rotation of the LRL vector with respect to a fixed reference direction has the smallest fluctuations. In order to increase the precision of calculations I have taken for the reference direction of each planet a direction close to the initial direction of the LRL vector. This is why the calculated average PA per century of each planet is slightly different from that obtained by in precedent works (\cite{Pogossian2022}, \cite{Pogossian2023}). In order to increase the time interval where the average of PA  is calculated I have used the entire time interval of 913 years.

To better understand Vulcan's influence on the orbits of inner planets, it is first necessary to analyze the partial perturbation induced by each planet in the solar system on the PA of each inner planet, since the PA of a single planet is sometimes subject to opposing contributions from other planets. So, before analyzing the influence of the hypothetical planet Vulcan on the inner planets, it's very useful and instructive to establish a table of the individual influence of each planet of our solar system on the inner planets.

So, in the preliminary calculations, I use a non-relativistic model of the solar system with eight planets and the Sun (9-body problem) and calculate a table of the individual contributions of the solar system planets to the PA of the inner planets.

Then I will add Vulcan to the used non-relativistic model of the solar system, and focus on Vulcan's influence on the PA of the inner planets. I will use two sets of parameters for orbital elements of Vulcan reported in a previous work (\cite{Pogossian2023}). 

In the first set of parameters, that  I call Le Verrier’s set,  I used the values of orbital elements for Vulcan reported by Le Verrier : \textit{a} = 0.1427 AU, e = 0, $ \textit i=12^\circ 10^{\prime}$ and $\Omega = 12^\circ 59^{\prime}$. For circular orbits argument of periapsis $\omega$ is not needed, and  the true anomaly $\nu$ of the Vulcan has been chosen arbitrarily equal to $45^\circ$.   Here \textit{a} is the semi-major axis, e is the eccentricity, $\textit{i}$ is the angle of inclination of the planet's orbital plane to the ecliptic and $\Omega$  is the longitude of the ascending node. As for Vulcan's mass, I have assigned it a mass equal to around 3.4 times that of Mercury, in order to induce an additional PA of Mercury  about $43^{\prime\prime}$/cy  in accordance with observation-based data.

The second set of parameters, which I call the optimized set, was calculated taking into account the limitations imposed by observational data and based on the principle of the most important influence of Vulcan on Mercury's orbit as described in Pogossian (2023) :   \textit{a} =0.18 AU, e=0.001, $\textit{i} = 7.0138675^\circ$,  	$\Omega = 48.1238724^\circ$, $\omega = 25.8162010^\circ$  and $\nu = 45^\circ$.  Using this optimized set of parameters, the mass of Vulcan was set to about 1.6 times the mass of Mercury in order to induce an additional PA of Mercury about  $43^{\prime\prime}$/cy  as described in \cite{Pogossian2023}.

\section{Results and Discussion}
\label{sec:3}
\
The PA of inner planets of solar system including Mercury are on average linear in time over a time interval of several tens of thousands of years. The PA of Mercury has a highly oscillatory character of variable amplitude, and it is only it’s average over a large time interval relative to it’s period of revolution that evolves linearly, as shown by (\cite{Narlikar1985}, \cite{Pogossian2022}). Random oscillations can reach amplitudes of magnitude even greater than the mean evolution. It is precisely this large-amplitude oscillatory character of the PA of inner planets which explains why their average value over 100 years is considered representative of their linear evolution. As for Mercury, this 100-year interval is long enough to allow Mercury to make 415 revolutions around the Sun, enabling us to calculate this average linear trend with acceptable precision, and small enough for this linear trend to emerge distinctly. As I demonstrated in a previous work (\cite{Pogossian2022}) the PA of Mercury can be defined in three different and distinct methods: as the rotation of the line joining the sun and the orbital position closest to the sun (extended PA), the rotation of the apsis line of the mean elliptical trajectory (geometric PA) and finally the rotation of the Laplace-Rungé-Lentz vector (PA defined by LRL). As shown in the latter work the PA defined by LRL vector translates this time evolution with smallest oscillations amplitude around the mean evolution. 

In the present work, the average PA of four inner planets over a time interval of around 913 years, measured by the rotation of the LRL vector, has been studied, taking into account the partial contribution of each planet of solar system.  Nevertheless, it should be emphasized that in order  to ensure an acceptable standard deviation for the mean linear PA of Mercury, time intervals of several hundred years will have to be considered (\cite{Pogossian2022}).

An average linear trend is observed for inner planets but not for outer planets. As an example, the behavior of Jupiter's perihelion is highly non-linear due to the mean motion orbital resonance 2:5 of Jupiter and Saturn, the "great inequality", and therefore the description of the perihelion advance of Jupiter by a linear trend is not relevant (\cite{Montenbruck1989}).

\begin{figure*}[h!t]
	\centering
	{\includegraphics[width=0.9\linewidth]{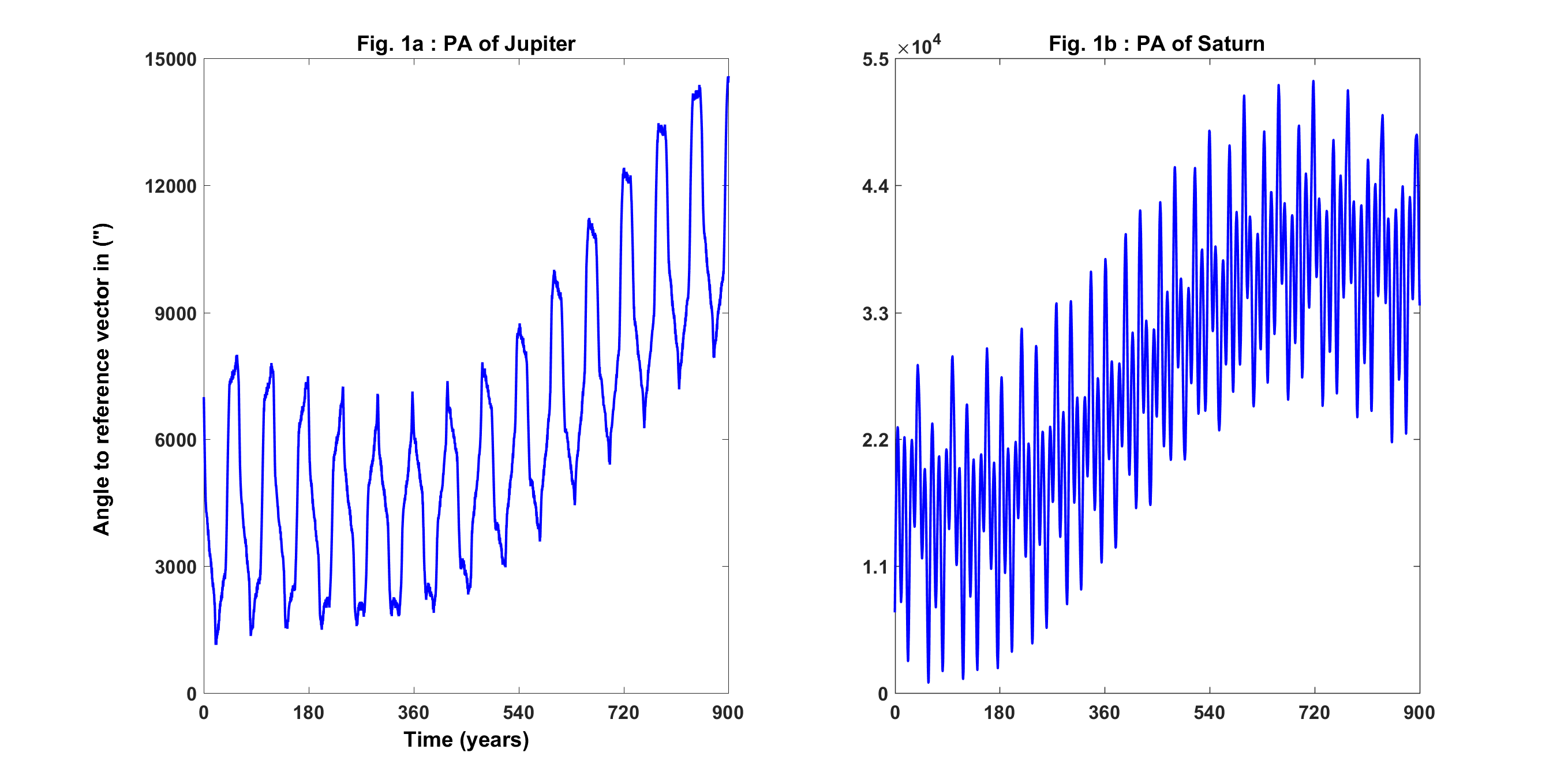}}
	\caption{In Fig. 1a, the evolution of Jupiter's PA over a time interval of 900 years is shown in angular seconds. The y-axis is off-set. A strong non-linear behavior is observed. Fig. 1b shows the evolution of Saturn's PA under the same conditions}
	\label{Fig1}
\end{figure*}
% Figure 1

Fig. 1 shows the highly non-linear behavior of the LRL vectors of Jupiter and Saturn, which are locked in an mean motion orbital resonance  2:5, and demonstrates that the average PA per century is inappropriate for describing the non-linear PA behavior of these two giant gaseous planets. The PAs of Uranus and Neptune under the influence of Saturn and Jupiter don't lend themselves to a linear function description either.

First, I'll calculate the distinct contribution of each planet in the solar system to the PA of Mercury, Venus, Earth and Mars, within the framework of Newton's classical theory of gravitation, calculated with Le Guyader's initial state vectors (\cite{Leguyader1993}) and without taking into account the hypothetical planet Vulcan.

Unlike the non-linear behavior of the PA of Jupiter and Saturn illustrated in On Fig. 1, the mean PA of Venus, Earth and Mars can be described by a linear function over a time interval of the order of 900 years, which is the time interval chosen for our work, as shown on Figs. 2a, 2b and 2c, respectively.

The evolution of the PA of Earth and Mars is less fluctuating over time than that of Venus, whose PA can undergo variations of up to $2500^{\prime\prime}$ in 10 years, whereas its average PA per century is only of the order of $34^{\prime\prime}$, as  Figs. 2a, 2b and 2c clearly show.

\begin{figure*}[h!t]
	\centering
	{\includegraphics[width=.9\linewidth]{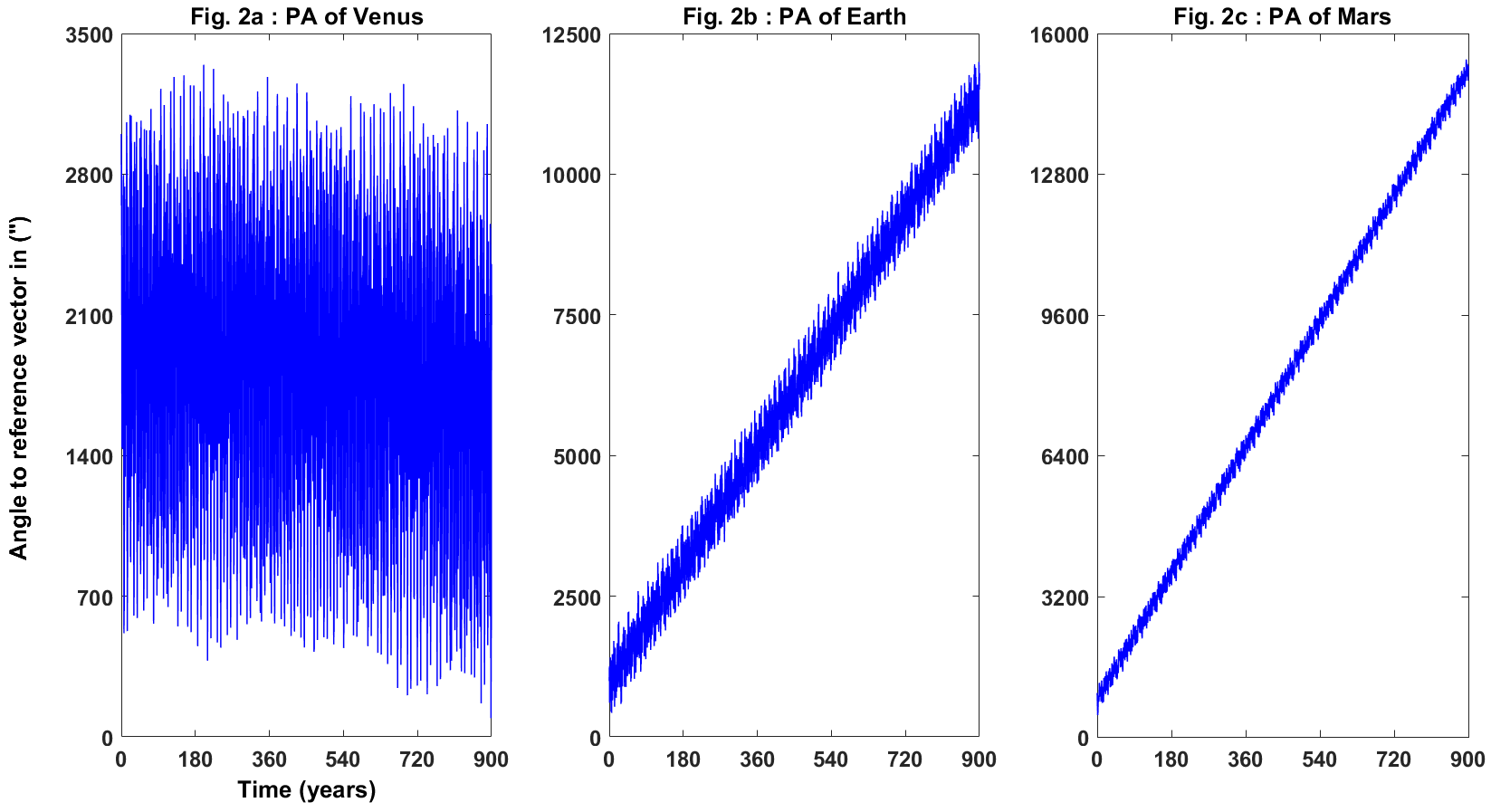}}
	\caption{In Fig. 2a, the mean linear trend of evolution of Venus’s PA over a time interval of 900 years is shown in angular seconds. The y-axis is off-set. Figs. 2b and 2c show respectively the evolution of  PA of Earth and Mars under the same conditions.}
	\label{Fig2} 
\end{figure*}
% Figure 2

Thus, the lowest accuracy in determining the PA of four inner planets is that of Venus. The large-amplitude oscillations in the PA of Venus are mainly due to the opposing contributions of the Earth and Jupiter.  To better understand this particular  behavior of the PA of Venus, the partial contributions of Earth and Jupiter to the PA of Venus will be studied separately.

The partial contribution of each solar system planet to the PA of Mercury, Venus, Earth and Mars has not been dealt with in depth in the past, and there are currently only a few works that address it  (\cite{Clemence1947}, \cite{Amelkin2019}, \cite{Biswas2008}). Although the method for evaluating these partial contributions is not presented in detail in the Clemence’s paper  (\cite{Clemence1947}), it is still the only comprehensive paper to provide numerical values for Mercury's  and Earth's PAs under the perturbing influences of other planets of solar system. In this respect, it's particularly important to understand the partial influence of each planet on the PA of the other planets in the solar system, as the PA of one planet can sometimes be subject to opposing contributions from other planets. In order to quantitatively account for the role and influence of each planet on the PA of the other planets, I analyze the orbital perturbation experienced by each planet of the solar system due to gravitational interaction with the other planets, by means of the rotation of the LRL vector, and by using two different methods.

The first method is straightforward. In the two-body problem, the perihelion of each planet remains fixed in space relative to the distant stars. A third body causes an orbital perturbation that result, inter alia, in the advance of the perihelion. First, I consider the motion of a single planet around the Sun, and then a second perturbing planet is introduced with Le Guyader's orbital parameters.  This method is therefore a study of the three-body problem within the framework of the Newtonian theory of classical gravitation, focusing on the evolution of the PA of each inner planet subjected to the gravitational perturbation of one of the other planets in the solar system.

To understand the unusual behavior of the Venus’s PA, I performed the following calculations using the first method. I calculated the average PA of Venus separately in the presence of Earth alone, Jupiter alone, then jointly with Earth and Jupiter respectively in Figs. 3a, 3b and 3c.

\begin{figure*}[h!t]
	\centering
	{\includegraphics[width=.9\linewidth]{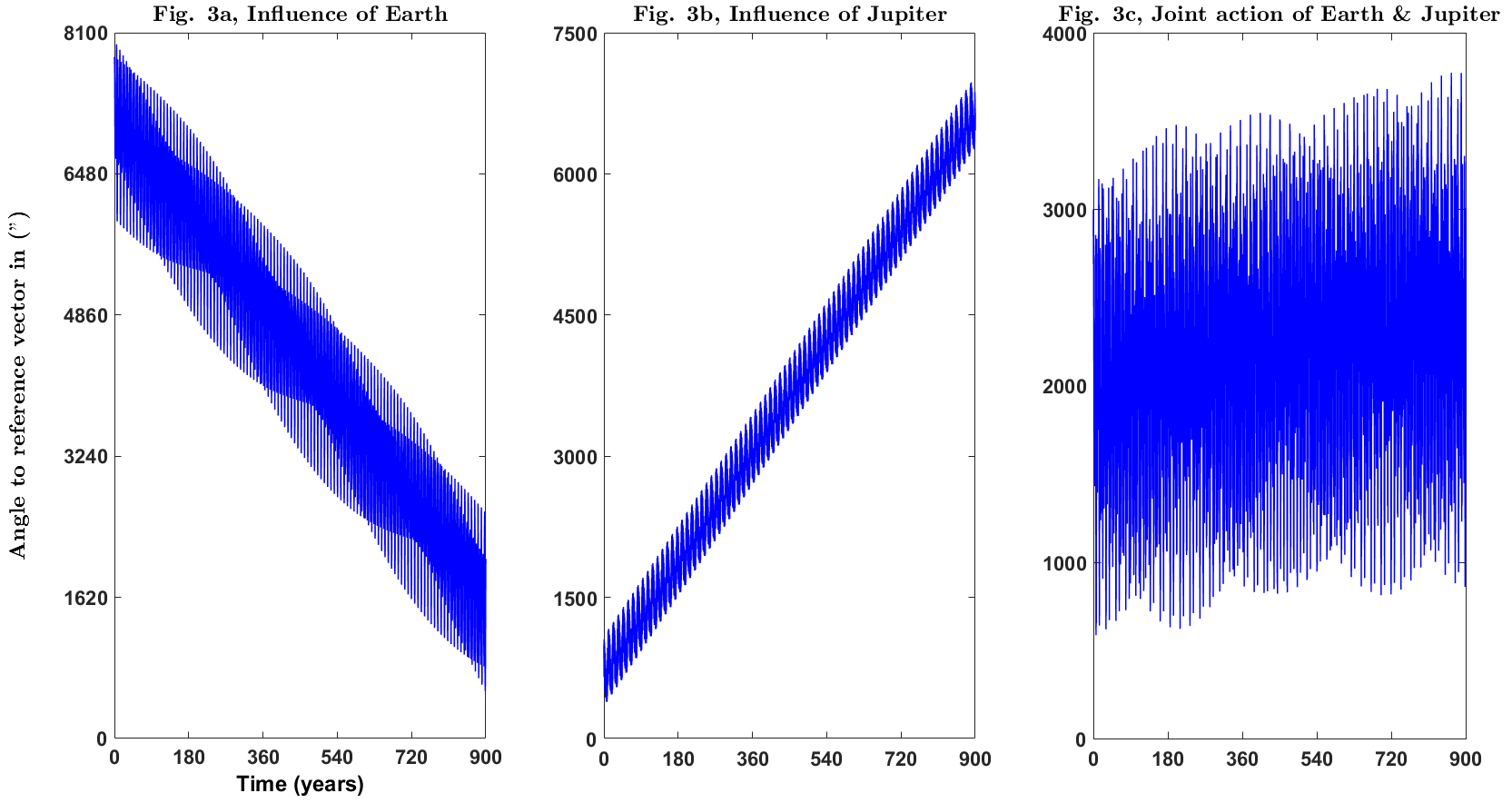}}
	\caption{In Fig. 3a, the linear trend of Venus' PA under the influence of the Earth over a 900-year time interval is shown in angular seconds. The y-axis is off-set. Fig. 3b shows the evolution of the PA of Venus in the presence of Jupiter alone under the same conditions. Fig. 3c shows the behavior of the PA of Venus under the joint influence of the Earth and Jupiter.}
	\label{Fig3} 
\end{figure*}
% Figure 3

The average PA per century of Venus under the influence of Jupiter alone is about $661^{\prime\prime}$, showing regular oscillations of $350^{\prime\prime}$ per 10 years around this linear trend. The average PA per century of Venus under the influence of Earth alone is around $611^{\prime\prime}$, but in the opposite direction, with fairly regular oscillations of $750^{\prime\prime}$ per 4 years.  The average PA per century of Venus under the joint influence of Earth and Jupiter is only $42^{\prime\prime}$ per century, with more irregular oscillations of $1200^{\prime\prime}$ per 4 years.

In the second method, I remove the disturbing planet from the list of planets in the N=9 body problem (or N=10 with Vulcan) and then study the influence of its absence on the PA of one of the inner planets. In this case, the original N=9-body problem (N=10 with Vulcan) is reduced to the N-1 body problem. In these calculations, the removal of one of the planets from the model is carried out by reducing its mass to 1 kg (negligible) while keeping the same initial orbital elements of all bodies. The difference between the PA of Mercury, Venus, Earth and Mars, calculated in the absence and presence of one of the planets in the solar system, can only be attributed, to a first approximation, to the perturbation of the removed planet.

Nevertheless, it is important to be aware of additional factors in this second method that contribute to the observed differences. For example, the presence or absence of Jupiter can significantly alter the orbits of neighboring planets, which in turn affects their individual gravitational contributions to the PA of any of the inner planets.

Using the second method, I have calculated in Fig. 4a the influence of the Earth on the PA of Venus, comparing the PA of Venus in the 9-body problem (all planets included) with the PA of Venus calculated without the Earth (8-body problem).  In Fig. 4b, the influence of Jupiter on the PA of Venus is calculated in the same way. 

\begin{figure*}[!htb]
	\centering
	{\includegraphics[width=.9\linewidth]{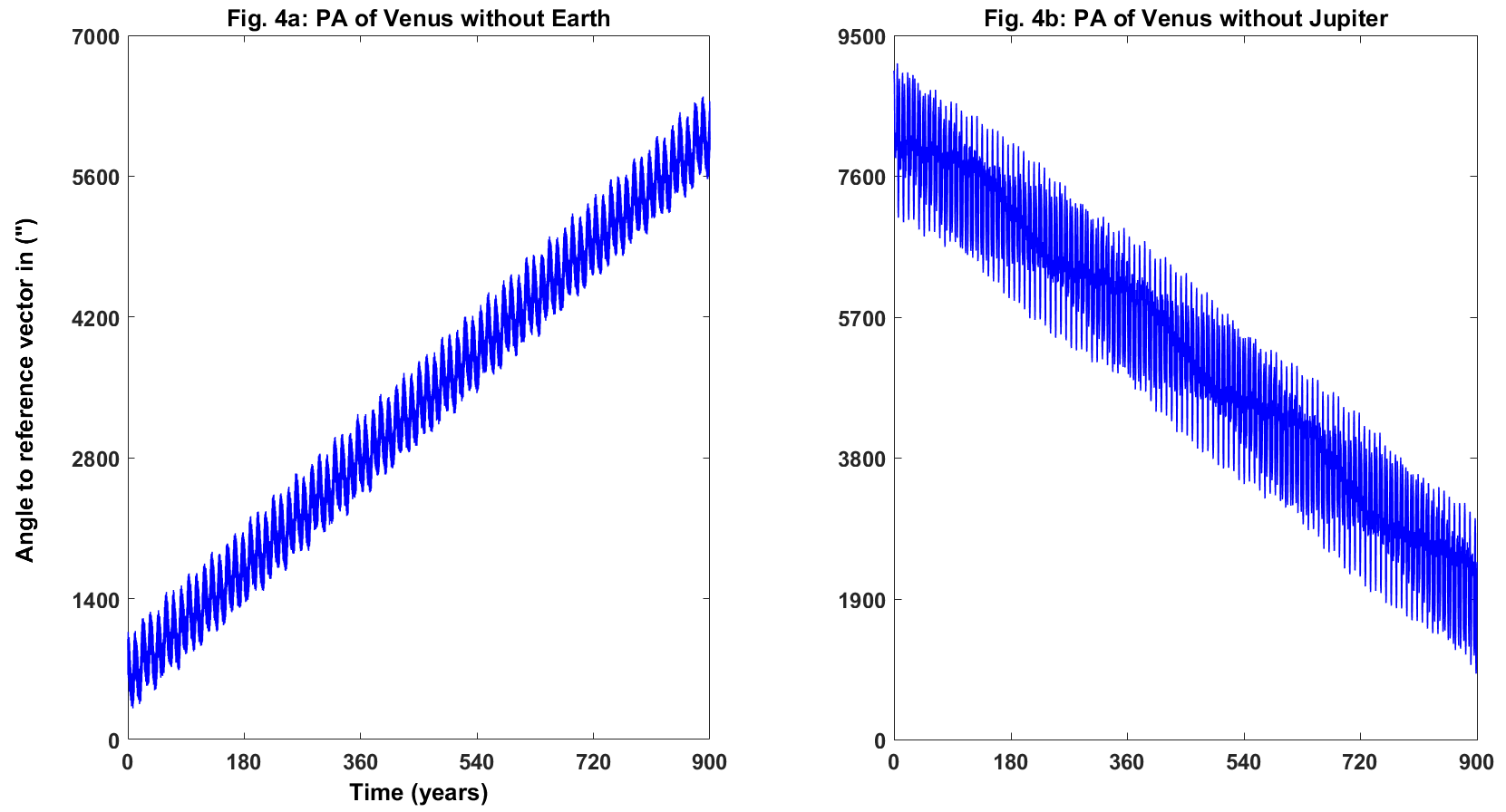}}
	\caption{The particularity of calculating the PA of Venus using the second method lies in the fact that, in the presence of all the other planets, we first make the Earth disappear in Fig. 4a, then reintegrate the Earth by making Jupiter disappear in Fig. 4b. In both cases, the PA of Venus is characterized by a pronounced linear trend, but in two different directions.}
	\label{Fig4} 
\end{figure*}

These results show that the behavior of Venus' PA over time under the influence of Earth alone and Jupiter alone is similar to the behavior calculated by the first method, with a slight difference in value ($659^{\prime\prime}$/cy for Jupiter and $-634^{\prime\prime}$/cy for Earth). The small difference in the Earth's behavior, calculated by the first and second methods may be explained by the fact that, in the second method, Jupiter modifies the Earth's orbit, so that the two calculations are not exactly identical, although they do reproduce the time dependence of the PA of Venus with satisfactory accuracy.

In Table \ref{tab:Table 1} and Table \ref{tab:Table 2}, the partial contributions of each planet in the solar system to the PA of the inner planets, calculated using the first and second methods, are presented.

The partial contribution of each planet to the PA of Mercury and the Earth are in very good agreement with the values presented by \cite{Clemence1947}. The small discrepancies between the Clemence’ reported values and my work may be attributed to differences in the initial values of orbital parameters and the non-relativistic model used by me.

\vskip12pt
\begin{table*}
%%	\tabularfont
	\caption{In this table the PA (in angular seconds per century: $^{\prime\prime}$/cy) of  inner planets, forming  each two body system with the sun, caused by a third perturbing planet added to a classical two body is given calculated by the first method. All PAs in this table are given in $^{\prime\prime}$/cy. I used  Le Guyader's orbital parameters (\cite{Leguyader1993})} 
	\label{tab:Table 1}
	\begin{center}
		\begin{tabular}{|p{1.2cm}||p{1.35cm}|p{1.3cm}|p{1.3cm}|p{1.3cm}|p{1.3cm}|p{1.3cm}|p{1.3cm}|p{1.3cm}|p{1.3cm}|} 
			\hline
			%				        &                                         \\ 
			\textbf{Planet} & \textbf{Summary \qquad \qquad influence} & {Mercury} & {Venus} & {Earth} & {Mars} & {Jupiter} & {Saturn} & {Uranus} & {Neptune}\\ 
			\hline
			Mercury & 531.96 & X & 277.23 & 90.83  & 2.48 & 153.78 & 7.22 & 0.14 & 0.04\\ 
			\hline
			Venus & -35.60 & -152.98 & X & -610.96 & 75.29 & 660.64 & 8.22 & 0.27 & 0.11\\
			\hline
			Earth & 1157.05 & -13.75 & 349.08 & X & 97.47 & 695.21 & 18.69 & 0.56 & 0.18\\
			\hline
			Mars & 1591.69 & 0.77 & 49.41 & 228.06 & X & 1248.16 & 65.57 & 1.19 & 0.34\\
			\hline
		\end{tabular}
	\end{center}
\end{table*}
\vskip12pt

As shown in Table \ref{tab:Table 1}, Mercury's PA is equal to $531.96^{\prime\prime}$cy in the N=9 body problem without Vulcan. The main contributions to the PA of Mercury are due to gravitational interactions with Venus ($277.23^{\prime\prime}$/cy) , Jupiter ($153.78^{\prime\prime}$/cy) and Earth ($90.83^{\prime\prime}$/cy), in agreement with the values presented by \cite{Clemence1947}.  In the case of the PA of Venus, the most significant contributions come from the perturbations caused by Venus' gravitational interaction with Jupiter ($661^{\prime\prime}$/cy), with Earth ($-611^{\prime\prime}$/cy),  with  Mercury ($-153^{\prime\prime}$/cy) and with Mars ($75^{\prime\prime}$/cy). Clearly, the greatest influence comes from the interaction of Venus with Earth and Jupiter, which contribute in opposite directions, considerably reducing the average PA per century. Also noteworthy are the contributions of Mercury and Mars in opposite directions. The PA of Venus amounts to $41.81^{\prime\prime}$/cy under the simultaneous influence of Earth and Jupiter (4-body problem), which is at least fifteen times less than the influence of each planet separately. This is also the reason why Venus' PA is subject to very strong oscillations, which considerably reduce the accuracy of the PA determination for this planet. 

The largest contributions to the Earth's PA come from the gravitational interaction with Jupiter ($695^{\prime\prime}$/cy), Venus ($349^{\prime\prime}$/cy) and Mars ($98^{\prime\prime}$/cy). These interactions contribute to the evolution in the same direction, which accounts for the high overall value ($1157^{\prime\prime}$/cy) of the Earth's PA, in agreement with the data reported by \cite{Clemence1947}. 

In Table \ref{tab:Table 2}, the PA of Mercury, Venus, Earth and Mars calculated by the second method are in agreement with those calculated by the first method (Table \ref{tab:Table 1}). To evaluate the PA induced by a planet, Saturn for example, on the inner planets, using the second method, the value of Saturn's mass is first reduced to 1 kg, while maintaining the same initial conditions in the simulator. The difference obtained in the PA of the inner planets, by comparing this calculation with another using Saturn's real mass, gives the partial contribution of  Saturn.

With the first method the influence of Saturn on the PA of Mercury, of Venus, of  Earth or of Mars are respectively $7.2^{\prime\prime}$/cy, $8.2^{\prime\prime}$/cy, $18.7^{\prime\prime}$/cy and $65.6^{\prime\prime}$/cy. Whereas with the second method, this influence amounts to $7.4^{\prime\prime}$/cy, $12.7^{\prime\prime}$/cy, $23.0^{\prime\prime}$/cy and $64.8^{\prime\prime}$/cy. 

As mentioned, the differences in the PA of the inner planets calculated by these two methods are due to the presence of Jupiter (and all other planets) in the second method, which strongly perturbs the orbits of all the planets as a function of their distance from Jupiter.

For various applications, the PA of inner planets with very low orbital inclinations is measured within the orbital plane relative to the direction of the equinox  $\dot{\varpi}=\dot{\omega}+cos\left( \textit{i}\right) ~\dot{\Omega}$  (\cite{Will2018}). The dot on the physical quantities signifies the derivative with respect to time. The initial values of these quantities can be estimated from the DE200 ephemeris  (\cite{Standish1990}). 
Unfortunately, I can't compare the perihelion advance rate measured via $\dot{\varpi}$ with my own calculations, as PA has been defined differently in my work.

\begin{table*}
	\caption{In this table, the PA (in angular seconds per century: $^{\prime\prime}$/cy) of inner planets is calculated in the absence of one of the solar system's perturbing planets (the absence is obtained by reducing the perturbing planet's size to 1 kg). The partial contribution of the perturbing planet is calculated (in brackets in the table) as the difference between the PA of one of the inner planets in the presence of all the planets and the PA of the same planet in the absence of one of the perturbing planets. All PAs in this table are given in $^{\prime\prime}$/cy. In the calculations I used the initial orbital elements of planets proposed by \cite{Leguyader1993}.} 
	\label{tab:Table 2}
	\begin{center}
		\begin{tabular}{|p{1.2cm}||p{1.4cm}|p{1.35cm}|p{1.2cm}|p{1.35cm}|p{1.2cm}|p{1.2cm}|p{1.2cm}|p{1.2cm}|p{1.2cm}|} 
			\hline
			%				        &                                         \\ 
			\textbf{Planet} & \textbf{Summary \qquad \qquad influence} & {Without \qquad Mercury} & {Without \qquad Venus} & {Without \qquad Earth} & {Without \qquad Mars} & {Without \qquad Jupiter} & {Without \qquad Saturn} & {Without \qquad Uranus} & {Without \qquad Neptune}\\ 
			\hline
			Mercury & 531.96 & X & 254.69 \qquad (277.27) & 441.12 \qquad (90.84)  & 529.47 \qquad (2.49) & 378.01 \qquad (153.95) & 524.86 \qquad (7.44) & 531.86 \qquad (0.10) & 531.92 \qquad (0.04)\\ 
			\hline
			Venus & -35.60 & 133.40 \qquad (-169.00) & X & 598.78 \qquad (-634.38) & -114.53 \qquad (78.93) & -694.87 \qquad (659.27) & -48.34 \qquad (12.74) & -35.78 \qquad (0.18) & -35.70 \qquad (0.10)\\
			\hline
			Earth & 1157.05 & 1170.50 \qquad (-13.45) & 803.17 \qquad (353.88) & X & 1058.39 \qquad (98.66) & 452.90 \qquad(704.15) & 1134.06 \qquad(22.99) & 1156.61 \qquad(0.44) & 1156.89 \qquad(0.16)\\
			\hline
			Mars & 1591.69 & 1590.97 \qquad(0.72) & 1542.44 \qquad (49.25) & 1364.50 \qquad(227.19) & X & 345.34 \qquad(1246.35) & 1526.85 \qquad (64.86) & 1590.50 \qquad(1.19) & 1591.35 \qquad (0.34)\\
			\hline
		\end{tabular}
	\end{center}
\end{table*}

Our calculations of the partial contribution of each planet in solar system to the PA of each inner planet, using proposed two different methods, are in good agreement with those provided by \cite{Clemence1947}. It is noteworthy also that the PA of the Earth and Mars show an average linear trend over the 900-year calculation time interval.  It is also worth mentioning other results on the partial contributions of solar system planets on the PA of inner planets (in \cite{Loskutov2011} one can find the PA of Mercury and Earth, in \cite{Amelkin2019} one can find the PA of Mercury), which give values consistent with those reported by \cite{Clemence1947} for the PA of Mercury and Earth.

From now on, I extend the solar system model to include the hypothetical planet Vulcan, thus solving the problem of N=10 bodies. The calculations in the presence of Vulcan are carried out for the two sets of orbital parameter values for the latter planet, the Le Verrier set and the optimized set of \cite{Pogossian2023} computed by the second method only , i.e., calculating the PA of one of the inner planets in the presence and absence of Vulcan, and then evaluating Vulcan's partial contribution by the difference between these two PA. This is relevant, all the more so as the values obtained for the PA of inner planets by the two methods show no significant differences.

The results of Vulcan's partial contribution to the PA for these planets is compared with the relativistic corrections provided by \cite{Pitjeva2010}. According to \cite{Pitjeva2010} the  GR suggests the following PA values for inner planets: $42.98^{\prime\prime}$/cy for Mercury, $8.62^{\prime\prime}$/cy for Venus, $3.84^{\prime\prime}$/cy for Earth, and $1.35^{\prime\prime}$/cy for Mars. Other authors reported PA values of inner planets due to relativistic effects (\cite{Biswas2008}, \cite{Capistrano2016}, \cite{Wilhelm2014}, \cite{Clemence1947}, \cite{Duncombe1956}) that are not significantly different from those reported by \cite{Pitjeva2010}. 

Let's first consider the case of Venus. Note that Venus' trajectory is very close to a circular orbit. For exactly circular orbits, the LRL vector and the perihelion are not defined. Furthermore, as underlined earlier in this work, the PA of Venus is the weakest compared to those of Mercury, Earth and Mars, with large oscillations of varying amplitude mainly due to the opposing contributions of Earth and Jupiter. More generally, the ambiguity of the notion of perihelion for the open trajectories of the planets of the solar system has been analyzed in one of my previous works [\cite{Pogossian2022}]. Therefore, the any comparison between the results for the PA of Venus obtained by theoretical calculations based on the GR, the Vulcan hypothesis and observational results are irrelevant.

As can be seen in Table \ref{tab:Table 3}, I obtain a PA due to the influence of Vulcan of $43.35^{\prime\prime}$/cy on Mercury, $3.57^{\prime\prime}$/cy on Venus, $1.17^{\prime\prime}$/cy on Earth and $0.27^{\prime\prime}$/cy on Mars for Le Verrier’ set of parameters for Vulcain. These results differ by at least a factor of 2 from those for Venus, Earth and Mars obtained by GR (\cite{Pitjeva2010}).  

\begin{table*}
	\caption{In this table, the partial contribution of Vulcan on the PA (in angular seconds per century) of  inner planets is calculated by the second method, i.e., by calculating the difference of PA of inner planets in  in the presence and in the absence of Vulcan. For the calculations, I used the initial orbital elements of the planets proposed by \cite{Leguyader1993} and as for the orbital elements of Vulcan, Le Verrier's set was used. In columns 5, 6, and 7 are reported  the calculated GR values for PA of the inner planets reported by \cite{Pitjeva2010},  by \cite{Clemence1947} and by \cite{Duncombe1956} respectively. All PAs in this table are given in $^{\prime\prime}$/cy.} 
	\label{tab:Table 3}
	%	\begin{center}
	\begin{tabular}{|p{1.5cm}||p{1.9cm}|>{\raggedright\arraybackslash}p{1.9cm}|p{1.9cm}|p{1.9cm}|p{1.9cm}|p{1.9cm}|} 
		\hline
		%				        &                                         \\ 
		\textbf{\qquad Planet} & \textbf{Influence of all planets} & {Influence of all planets and Vulcan} & {Influence \qquad of Vulcan} &  {\cite{Pitjeva2010}} & {\cite{Clemence1947}} & {\cite{Duncombe1956}}\\ 
		\hline
		Mercury & 531.96 & 575.31 & 43.35 & 42.98 & 43.50 & 43.03 \\ 
		\hline
		Venus & -35.60 & -32.03 & 3.57 & 8.62 & 8.62 & 8.6\\
		\hline
		Earth & 1157.05 & 1158.22 & 1.17 & 3.84 & 3.87 & 3.8\\
		\hline
		Mars & 1591.69 & 1591.97 & 0.27 & 1.35 & 1.36 & X \\
		\hline
	\end{tabular}
	%	\end{center}
\end{table*}

The PA induced by Vulcan calculated using the optimized set of Vulcan’s orbital elements is even more divergent from the corresponding values obtained by GR, presented in Table \ref{tab:Table 4}. These results differ by  more than factor of 2 from those for Venus, Earth and Mars obtained by GR (\cite{Pitjeva2010}).

\begin{table*}
	\caption{In this table, the partial contribution of Vulcan on the PA (in angular seconds per century) of  inner planets is calculated by the second method, i.e. , by calculating the difference of PA of inner planets in  in the presence and in the absence of Vulcan. All PAs in this table are given in $^{\prime\prime}$/cy. In the calculations I used the initial orbital elements of planets proposed by \cite{Leguyader1993}. In this table for orbital elements of Vulcan the optimized set was used.} 
	\label{tab:Table 4}
	\begin{center}
		\begin{tabular}{|p{1.3cm}||p{1.8cm}|>{\raggedright\arraybackslash}p{1.8cm}|p{1.8cm}|} 
			\hline
			%				        &                                         \\ 
			\textbf{\qquad Planet} & \textbf{Influence of all planets} & {Influence of all planets and Vulcan} & {Influence of Vulcan}\\ 
			\hline
			Mercury & 531.96 & 575.31 & 43.35  \\ 
			\hline
			Venus & -35.60 & -32.86 & 2.75 \\
			\hline
			Earth & 1157.05 & 1157.95 & 0.88 \\
			\hline
			Mars & 1591.69 & 1591.89 & 0.20  \\
			\hline
		\end{tabular}
	\end{center}
\end{table*}

Thus the comparison of reported observation values, the values obtained from GR correction to PA of inner planets, estimated by different authors,  points to a difference of at least a factor of 2 with the correction provided by the Vulcan hypothesis for 2 sets of orbital elements capable of ensuring a value of $43^{\prime\prime}$/cy for Mercury's PA.

To understand these results in more detail, let's take a careful examination of how the anomalous PA of the inner planets are determined from observations. 

In the most recent ephemerides (\cite{Park2017}, \cite{Park2021}, \cite{Newhall1983}, \cite{Standish1990}) a large number of astrometric position measurements of various types, including optical data, high-precision radar ranging, spacecraft ranging and lunar laser ranging, are used to determine over several hundred parameters of planets, asteroids, the Sun's standard gravitational parameter, astronomical unit, etc., from a least-squares fit using a parametrized post Newtonian N body metric for the solar system. The uncertainties of the determination of these parameters decrease with increasing the time interval of observations.

These data are then used to extract anomalous PA using models based mainly on GR.  Published data provide little information on the definition of perihelion, and even less on the time interval used to determine the mean linear trend required in the PA calculations. As shown in one of my previous works (\cite{Pogossian2022}), the linear trend, which gives the average value of PA of the inner planets, can only be obtained with acceptable accuracy over a long time interval of at least several hundred years due to large-amplitude oscillations of the average PA value for three different definitions of Mercury's perihelion.  One of the three PA definitions based on the use of the Laplace-Runge-Lentz rotation gives the lowest sensitivity with respect to the averaging time interval, which however requires at least several hundred years to ensure acceptable accuracy.

Thus, the value of PA considered as observational, based on a database of only a few decades without giving the exact definition of perihelion, the model used to extract its value, and the averaging time interval to determine the linear trend cannot serve as a reference for comparison my calculated values. 

As for the comparison of theoretical quantities between different models, they must have identical definitions, otherwise their comparison remains either very vague or impossible.

With the interrogations I've expressed about what is considered the observational value of PA, my results allow me to conclude that Vulcan's PBH hypothesis as a possible explanation for Mercury's AP anomaly does not give the same PA values for Venus, Earth and Mars, as those attributed to GR.  The difference between the values I have reported and those attributed to GR may be due to the fact that the Vulcan hypothesis is not relevant. 

This difference may also be due to the fact that the PA calculated in GR and by rotation of the LRL vector used in this work are not defined in the same way. 

Finally, it should be noted that the inclusion of Vulcan also affects other orbital characteristics. To illustrate this, let's examine how the aphelion distances of the inner planets from the Sun are influenced by Vulcan during their 101st revolution, using two different solar system models. One 10-body solar system model includes Vulcan, while the other, with 9 bodies, omits it. The difference between these distances is scaled down to Mercury's radius, to better reflect the extent to which this difference can be detected in astronomical measurements.

\begin{table*}
	\caption{This table shows the differences in distance to aphelion for each planet during the 101-th revolution, with and without Vulcan. These values are given at Mercury's radius scale and are specified for two sets of parameters : the optimized set and Le Verrier's parameters.}
	
	\label{tab:Table 5}
	\begin{center}
		\begin{tabular}{|>{\raggedright\arraybackslash}p{2.5cm}||p{1.9cm}|p{1.9cm}|p{1.9cm}|p{1.9cm}|} 
			\hline
			%				        &                                         \\ 
			Differences are scaled with $1.6325\times 10^{-5} AU$ & {Difference for	Mercury} & {Difference for Venus} & {Difference for Earth} & {Difference for Mars}\\ 
			\hline
			Optimized set & 0.1615 & 0.6039 & -1.1073   & 1.4515 \\ 
			\hline
			Le Verrier's set & 0.2149 & 0.5910 & -1.0135  & 1.6209 \\
			\hline
		\end{tabular}
	\end{center}
\end{table*}

Hundred orbital revolutions of Mercury, Venus, Earth and Mars correspond to time intervals of about 24, 61.5, 100 and 188 years, respectively. In particular, the 188-year interval for the 100 revolutions of Mars is particularly well suited to observing cumulative deviations over longer time scales. This may partly explain the increased differences observed between models with and without Vulcan for planets with higher periods of revolution.  It is important to observe that Table \ref{tab:Table 5}  reveals significant differences, including a negative value specifically for Earth, that are readily detectable using current observational methods.

In conclusion, my work does not definitively rule out the Vulcan hypothesis as a possible explanation for the anomalous displacement of Mercury's perihelion, but it would require to be confirmed or invalidated by long-term observations.

\vspace{2em}

\section{Conclusion}

The PBH Vulcan hypothesis could explain the anomalous advance of Mercury's perihelion caused by its gravitational attraction, which has remained hidden from astronomers. In this case, its gravitational interaction with the other inner planets (Venus, Earth and Mars) will lead to an additional advance in the perihelion of the latter. 

I carry out a careful analysis of the partial PA of each inner planet under the influence of gravitation pull of other planets of solar system, including Vulcan, within the framework of the Newtonian theory of classical gravitation.  To this goal, two different methods are proposed, the first based on the three-body problem and the second by cancelling out the mass of the influencing planet in N (N=9,10) body calculations.

One of the noteworthy results of this analysis is that it revealed the low accuracy of the determination of Venus' perihelion advance. The large-amplitude chaotic oscillations in Venus' perihelion advance are mainly due to the opposing contributions of the Earth and Jupiter. So, the comparison between the calculations based on GR, the results obtained using the Vulcan hypothesis and the observational data is irrelevant for Venus. 

Due to the highly non-linear PA behavior of the outer planets over a 900-year period, my study has been limited to the inner planets. When including Vulcan into calculations two different set of values were used for its orbital parameters defined in a previous work on Vulcan (\cite{Pogossian2023}). 

The reliability of my calculations using two different methods was confirmed by comparing the partial contribution of each planet to the PA of Mercury and the Earth with the values presented by \cite{Clemence1947}.
My obtained values for  PA of Venus, Earth and Mars due to their interaction with Vulcan for the Le Verrier’s set of  Vulcan orbital parameters differ by at least a factor of 2 from those obtained by GR according to \cite{Pitjeva2010}. 
Those obtained for the optimized set of Vulcan orbital parameters show an even greater difference with GR data. 

For comparisons between different models and with observational data, it's important to clearly define the perihelion for the real orbits of the inner planets.

Can we conclude from this that the Vulcan hypothesis can now be dismissed as a reliable explanation for the anomalous PA of  Mercury ? 

The present work does not definitively rule out the Vulcan hypothesis as a possible explanation for the anomalous advance of Mercury's perihelion, but it would need to be confirmed or invalidated by long-term observations. To do this, the data used to determine the mean PA of inner planets from observations must be collected over at least several hundred years, due to the large-amplitude oscillations in the mean PA value of the inner planets. In addition, the theoretical model used for the observational data reduction should be carefully selected.

\begin{acknowledgments}
The author wishes to express his sincere gratitude to X. Carton, A. Colin de Verdière, G. Roullet and A. Pogosyan for their valuable discussions. Their contributions and expertise have helped refine some of the ideas presented in this article.
\vspace{-1em}
\end{acknowledgments}

\clearpage
\bibliography{ArXiv_Pogossian_And_if_Vulcan_was_a_PBH.bib}{}
\bibliographystyle{aasjournal}

\end{document}